# A Digital Signature with Threshold Generation and Verification

Sunder Lal [*] and Manoj Kumar [**]

[*] *Dept of Mathematics, IBS Khandari.Dr. B.R.A.University Agra.*
Sunder_lal2@rediffmail.com.in.
[**] *Dept of Mathematics, HCST, Farah – Mathura,* (U. P.) – 281122.
Yamu_balyan@yahoo.co.in

**Abstract -** This paper proposes a signature scheme where the signatures are generated by the cooperation of a number of people from a given group of senders and the signatures are verified by a certain number of people from the group of recipients. Shamir's threshold scheme and Schnorr's signature scheme are used to realize the proposed scheme.

## 1. Introduction

Physical signature is a natural tool to **authenticate** the communication, but it is useless in electronic messages; one has to rely on other methods like **digital signature**. Digital signature is a cryptographic tool to solve this problem of authenticity in electronic communications. Basically digital signature has a property that anyone having a copy of the signature can check its validity by using some public-information, but no one else can forge the signature on another document. This property of digital signature is called **self-authentication.** In most situations, the signer and the verifier is generally a single person. However when the message is sent by one organization to another organization, a valid message may require the approval or consent of several people. In this case, the signature generation and verification is done by more than one consenting rather than by a single person. A common example of this policy is a large bank transaction, which requires the signature of more than one person. Such a policy could be implemented by having a separate digital signature for every required signer, but this solution increases the effort to verify the message linearly with the number of signer. **Threshold signature** is an answer to this problem.

The ($t$ , $n$) threshold signature schemes [1,2,3,5,6,8] are used to solve these problems. Threshold signatures are closely related to the concept of threshold cryptography, first introduced by Desmedt [1,2]. In 1991 Desmedt and Frankel [1] proposed the first (t, n) threshold digital signature scheme based on the RSA assumption.

This paper proposes a **Threshold signature scheme with threshold verification** based on Shamir's threshold signature scheme [10] and Schnorr's signature scheme [9]. These basic tools are briefly described in the next section.

The section-2 presents **some basic tools**. In Section-3 we present a **Threshold Signature Scheme with threshold verification**. Section-4 discusses the **security of the Scheme**. An **illustration** to the scheme is discussed in section-5. **Remarks** are in section-6.

## 2. Preliminaries

Throughout this paper we will use the following system setting.

- A prime modulous $p$, where $2^{511} < p < 2^{512}$;
- A prime modulous $q$, where $2^{159} < q < 2^{160}$ and $q$ is a divisor of $p - 1$;
- A number $g$, where $g \equiv k^{(p-1)/q} \bmod p$, $k$ is random integer with $1 \leq k \leq p-1$ such that g >1; (g is a generator of order $q$ in $Zp^*$).
- A collision free one-way hash function $h$ [12];

The parameters $p, q, g$ and $h$ are common to all users. Every user has two keys one private and one public. We assume that a user A chooses a random $x_A \in Zq$ and computes $y_A = g^{x_A} \bmod p$. He keeps $x_A$ as his private key and publishes $y_A$ as his public key.

In **Schnorr's signature scheme** the signature of the user A on the message $m$ is given by $(r_A, S_A)$, where, $r_A = h(g^{k_A} \bmod p, m)$, and $S_A = k_A - x_A \cdot r_A \bmod p$. The signature are verified by checking the equality $r_A = h(g^{S_A} y^{r_A} \bmod p, m)$.

A **(t, n) threshold secret sharing scheme** is a scheme to distribute a secret key K into $n$ users in such a way that any $t$ users can cooperate to reconstruct K but a collusion of $t - 1$ or less users reveal nothing about the secret. There are many realization of this scheme, we shall use Shamir's scheme. This scheme is based on Lagrange interpolation in a field. To implement it, a polynomial $f$ of degree $t - 1$ is randomly chosen in $Zq$ such that $f(0) = K$. Each user $i$ is given a public identity $u_i$ and a secret share $f(u_i)$. Now any subset of $t$ shareholders out of $n$ shareholders can reconstruct the secret K = $f(0)$, by pooling their shares and using

$$f(0) = \sum_{i=1}^{t} f(u_i) \prod_{j=1, j \neq i}^{t} \frac{-u_j}{u_i - u_j} \bmod q$$

Here for simplicity the authorized subset of $t$ users consists of shareholders $i$ for $i = 1,2,3\ldots t$.

## 3. The proposed scheme.

The signer of the conventional digital signature schemes is usually a single person. But when the message is transmitted by an organization $S$ to another organization $R$ and may require the approval of more than one person then the responsibility of signing the messages needs to be shared by a subset $H_S$ of $t$ or more signer from a designated group $G_S$ of $n$ users belongs to the organization $S$.

On the other hand, the signing group wants to generate the signature on a message $m$ in such a way that the signature can be verified by any subset $H_R$ of $k$ or more users from a designated group $G_R$ of $l$ users belongs to the organization $R$, then threshold verification schemes serve our purpose. This paper proposes a threshold signature scheme with threshold verification.

We assume that the secret key of the organization $S$ is $x_S$ and the public key is $y_S$ where $y_S = g^{x_S}$ mod $p$ with $x_S \in Z_q$. Similarly the organization $R$ possesses a pair $(x_R, y_R)$ of private key $x_R$ and public key $y_R = g^{x_R}$ mod $p$. Also every user A in both the organizations possesses a pair $(x_A, y_A)$ with $x_A$ secret and $y_A = g^{x_A}$ mod $p$ public.

We further assume that both the organization $R$ and $S$ have a common trusted center (CTC) for determining the group secret parameters of the two groups and also the secret shares all members. This scheme consists of the following steps: -

### 3.1. Group Secret Key and Secret Shares Generation for the organization S.

(a). CTC selects the group public parameters $p$, $q$, $g$ and a collision free one way hash function $h$. CTC also selects for the group $G_S$ a polynomial

$$f_S(x) = a_0 + a_1 x + \ldots a_{t-1} x^{t-1} \bmod q, \text{ with } a_0 = x_S = f_S(0).$$

(b). CTC computes the group public key, $y_S$, as, $y_S = g^{f_S(0)} \bmod p$.

(c). CTC randomly selects $K \in Z_q$ and computes a public value $W = g^{-K} \bmod p$.

(d). CTC computes a public value $v_{S_i}$ for each member of the group $G_S$, as,

$$v_{S_i} = f_S(u_{S_i}) \cdot y_{S_i}^K \bmod p.$$

Here, $y_{S_i}$ is public key and $u_{S_i}$ is the public value associated with each user $i$ in the group $G_S$.

(e). CTC sends $\{v_{S_i}, W\}$ to each user $i$ in the group $G_S$ through a public channel.

### 3.2. Group Secret Key and Secret Shares Generation for the organization R

(a). CTC selects for the group $G_R$ a polynomial

$$f_R(x) = b_0 + b_1 x + \ldots b_{k-1} x^{k-1} \bmod q \text{ with } b_0 = x_R = f_R(0).$$

(b). CTC computes the group public key, $y_R$, as, $y_R = g^{f_R(0)} \bmod p$.

(c). CTC computes a public value $v_{R_i}$ for each member of the group $G_R$, as,

$$v_{R_i} = f_R(u_{R_i}) \cdot y_{R_i}^K \bmod p.$$

Here, $y_{R_i}$ is public key and $u_{R_i}$ is the public value associated with each user $i$ in the group $G_R$.

(d). CTC sends $\{v_{R_i}, W\}$ to each user $i$ in the group $G_R$ through a public channel.

### 3.3. Signature generation by any t users

If a $H_S$ is subset of $t$ members of the organization $S$ out of $n$ members who agree to sign a message $m$ to be sent to the organization $R$, then the signature generation has the following steps.

(a). Each user $i \in H_S$ randomly selects $K_{i_1}$ and $K_{i_2} \in Z_q$ and computes

$$u_i = g^{-K_{i_2}} \bmod p, \quad v_i = g^{K_{i_1}} \bmod p \text{ and } w_i = g^{K_{i_1}} y_R^{K_{i_2}} \bmod p.$$

(b). Each user $i$ broadcasts $u_i, w_i$ publicly and $v_i$ secretly to every other user in $H_S$. Once all $u_i, v_i$ and $w_i$ are available, each member $i, i \in H_S$ computes the product $U_S, V_S, W_S$ and a hash value $R_S$, as,

$$U_S = \prod_{i \in H_S} u_i \bmod q, \quad V_S = \prod_{i \in H_S} v_i \bmod q,$$

$$W_S = \prod_{i \in H_S} w_i \bmod q \text{ and } R_S = h(V_S, m) \bmod q.$$

(c). Each user $i \in H_S$ recovers his/her secret share $f_S(u_{S_i})$, as, $f_S(u_{S_i}) = v_{S_i} W^{x_{S_i}} \bmod p$.

(d). Each user $i \in H_S$ modifies his/her shadow, as, $MS_{S_i} = f_S(u_{S_i}) \cdot \prod_{j=1, j \neq i}^{t} \frac{-u_{S_j}}{u_{S_i} - u_{S_j}} \bmod q$.

(e). By using his/her modified shadow $MS_{S_i}$, each user $i \in H_S$ computes his/her partial signature $s_i$, as, $s_i = K_{i_1} + MS_{S_i} \cdot R_S \bmod q$.

(f). Each user $i \in H_S$ sends his/her partial signature to the CTC, who produce a group signature $S_S$, as, $S_S = \sum_{i=1}^{t} s_i \bmod q$.

(g). CTC sends $\{S_S, U_S, W_S, m\}$ to the **designated combiner DC of organization R** as signature of the group $S$ for the message $m$.

### 3.4. Signature verification by the organization R

Any subset $H_R$ of $k$ users from a designated group $G_R$ can verify the signature. We assume that there is **designated combiner DC (can be any member among the members of the group $G_R$, or** the head of the organization *R*), who collects partial computations from each user in $H_R$ and determines the validity of signature. The verifying process is as follows.

(a) Each user $i \in H_R$ recovers his/her secret share, as, $f_R(u_{R_i}) = v_{R_i} W^{x_{R_i}} \mod p$.

(b) Each user $i \in H_R$ modifies his/her shadow, as, $MS_{R_i} = f_R(u_{R_i}) \cdot \prod_{j=1, j \neq i}^{k} \frac{-u_{R_j}}{u_{R_i} - u_{R_j}} \mod q$.

(c) Each user $i \in H_R$ sends his modified shadow $MS_{R_i}$ to the *DC*. *DC* computes $R_R$, as,

$$R_R = W_S . U_S^{\sum_{i=1}^{k} MS_{R_i}} \mod q \text{ and recovers } R_S = h(R_R, m) \mod q.$$

(d). *DC* checks the following congruence $g^{S_S} \stackrel{?}{\equiv} R_R \cdot y_S^{R_S} \mod p$ for a valid signature.

If hold then $\{S_S, U_S, W_S, m\}$ is a valid signature on the message *m*.

**4. Security discussions.**

In this sub-section, we shall discuss the security aspects of proposed **Threshold Signature Scheme with threshold verification.** Here we shall discuss several possible attacks. But none of these can successfully break our system.

(a). Can any one retrieve the organization's secret keys $x_S = f_S(0)$ and $x_R = f_R(0)$, from the group public key $y_S$ and $y_R$ respectively?

This is as difficult as solving discrete logarithm problem. No one can get the secret key $x_S$ and $x_R$, since $f_S$ and $f_R$ are the randomly and secretly selected polynomials by the CTC. On the other hand, by using the public keys $y_S$ and $y_R$ no one also get the secret keys $x_S$ and $x_R$ because this is as difficult as solving discrete logarithm problem.

(b). Can one retrieve the secret shares, $f_S(u_{S_i})$ of members of $G_S$, from the equation

$$v_{S_i} = f_S(u_{S_i}) \cdot y_{S_i}^{K} \mod p ?$$

No because $f_S$ is a randomly and secretly selected polynomial and *K* is also a randomly and secretly selected integer by the CTC. Similarly no one can get the secret shares $f_R(u_{R_i})$ of members of $G_R$, from the equation $v_{R_i} = f_R(u_{R_i}) \cdot y_{R_i}^{K} \mod p$.

(c). Can one retrieve the secret shares, $f_S(u_{S_i})$ of members of $G_S$, from the equation

$$f_S(u_{S_i}) = v_{S_i} W^{x_{S_i}} \mod p ?$$

Only the user $i$ can recovers his secret shares, $f_S(u_{S_i})$, because $f_S$ is a randomly and secretly selected polynomial and $x_{S_i}$ is secret key of the user $i \in G_S$. Similarly no one can get the secret shares $f_R(u_{R_i})$ of members of $G_R$, from the equation $f_R(u_{R_i}) = v_{R_i} W^{x_{R_i}} \mod p$.

**(d).** Can one retrieve the modified shadow $MS_{S_i}$, integer $K_{i_1}$, the hash value $R_S$ and partial signature $s_i$, $i \in G_S$ from the equation $s_i = K_{i_1} + MS_{S_i} . R_S \mod q$ ?

They all are secret parameters and it is computationally infeasible for a forger to collect the $MS_{S_i}$, integer $K_{i_1}$, the hash value $R_S$ and partial signature $s_i$, $i \in G_S$.

**(e).** Can the designated CTC retrieve any partial information from the equation,

$$S_S = \sum_{i=1,}^{t} s_i \mod q \ ?$$

Obviously, it would be computationally infeasible for CTC to derive any information from $s_i$.

**(f).** Can one impersonate a member $i$, $i \in H_S$ ?

A forger may try to impersonate a shareholder $i$, $i \in H_S$, by randomly selecting two integers $K_{i_1}$ and $K_{i_2} \in Z_q$ and broadcasting $u_i$, $v_i$ and $w_i$. But without knowing the secret shares, $f_S(u_{S_i})$, it is difficult to generate a valid partial signature $s_i$ to satisfy the verification equation,

$$S_S = \sum_{i=1,}^{t} s_i \mod q \quad \text{and} \quad g^{S_S} \stackrel{?}{\equiv} R_{R.} \ y_S^{R_S} \mod p.$$

**(g).** Can one forge a signature $\{S_S, U_S, W_S, m\}$ by the following equation,

$$g^{S_S} \equiv R_{R.} \ y_S^{R_S} \mod p ?$$

A forger may randomly selects an integer $R_R$ and then computes the hash value $R_S$ such that $R_S = h(R_R, m) \mod q$.

Obviously, to computes the integer $S_S$ is equivalent to solving the discrete logarithm problem. On the other hand, the forger can randomly select $R_R$ and $S_S$ first and then try to determine a value $R_S$, that satisfy the equation $g^{S_S} \equiv R_{R.} \ y_S^{R_S} \mod p$.

However, according to the one-way property of the hash function $h$, it is quite impossible. Thus, this attack will not be successful.

**(h).** Can $t$ or more shareholders act in collusion to reconstruct the polynomial $f_S(x)$ ?

According to the equation $f_S(x) = \sum_{i=1}^{t} f_S(u_{S_i}) \prod_{j=1, j\neq i}^{t} \frac{x - u_{S_j}}{u_{S_i} - u_{S_j}} \mod q$, the secret polynomial $f_S(x)$ can be reconstructed with the knowledge of any $t$ secret shares, $f_S(u_{S_i})$, $i \in G_S$. So if in an organization the shareholders are known to each other, any $t$ of them may collude and find the secret polynomial $f_S$. This attack, however, does not weaken the security of our scheme in the sense that the number of users that have to collude in order to forge the signature is not smaller than the threshold value.

## 5. Illustration

We now give an illustration in support of our scheme. Suppose $|G_S| = 7$, $|H_S| = 4$, $|G_R| = 6$ and $|H_S| = 5$. We take the parameters $p = 47$, $q = 23$ and $g = 2$.

### 5.1. Group Secret Key and Secret Shares Generation for the organization S

(a). CTC selects for the group $G_S$, a polynomial $f_S(x) = 11 + 3x + 13x^2 + x^3 \mod 23$. Here $x_S = 11$ and $y_S = 2^{11} \mod 47 = 27$.

(b). CTC randomly selects $K = 8$ and computes $W = 2^{-8} \mod 47 = 9$.

(c). CTC computes the secret key, public key, secret share and public value for each member of the group $G_S$, as shown by the following table.

| VALUE / USER | $y_{S_i}$ | $x_{S_i}$ | $u_{S_i}$ | $f_S(u_{S_i})$ | $v_{S_i}$ |
|---|---|---|---|---|---|
| User – $S_1$ | 7 | 12 | 2 | 8 | 34 |
| User – $S_2$ | 37 | 10 | 9 | 3 | 34 |
| User – $S_3$ | 28 | 14 | 8 | 22 | 38 |
| User – $S_4$ | 18 | 16 | 11 | 4 | 9 |
| User – $S_5$ | 8 | 21 | 3 | 3 | 6 |
| User – $S_6$ | 3 | 19 | 5 | 16 | 6 |
| User – $S_7$ | 36 | 17 | 4 | 19 | 40 |

### 5.2. Group Secret Key and Secret Shares Generation for the organization R

(a). CTC selects for the group $G_R$ a polynomial $f_R(x) = 7 + 2x + 4x^2 + 3x^3 \mod 23$. Here $x_R = 7$ and $y_R = 2^7 \mod 47 = 34$.

(b). CTC computes the secret key, public key, secret share and public value for each member of the group $G_R$, as shown by the next table.

| VALUE USER | $y_{R_i}$ | $x_{R_i}$ | $u_{R_i}$ | $f_R(u_{R_i})$ | $v_{R_i}$ |
|---|---|---|---|---|---|
| User – $R_1$ | 9 | 15 | 11 | 21 | 14 |
| User – $R_2$ | 42 | 9 | 5 | 9 | 25 |
| User – $R_3$ | 27 | 11 | 8 | 21 | 16 |
| User – $R_4$ | 25 | 18 | 3 | 15 | 20 |
| User – $R_5$ | 17 | 6 | 6 | 6 | 24 |
| User – $R_6$ | 14 | 13 | 4 | 18 | 32 |

*5.3. Signature generation by any t users*

    If $S_2$, $S_4$, $S_6$ and $S_7$ are agree to sign a message *m* for the organization *R* , then

(a). $S_2$ selects $K_{2_1} = 5$, $K_{2_2} = 7$ and computes $u_2 = 18$, $v_2 = 32$ and $w_2 = 21$.

(b). $S_4$ selects $K_{4_1} = 4$, $K_{4_2} = 3$ and computes $u_4 = 6$, $v_4 = 16$ and $w_4 = 4$.

(c). $S_6$ selects $K_{6_1} = 12$, $K_{6_2} = 18$ and computes $u_6 = 32$, $v_6 = 7$ and $w_6 = 1$.

(d). $S_7$ selects $K_{7_1} = 21$, $K_{7_2} = 11$ and computes $u_7 = 7$, $v_7 = 12$ and $w_7 = 17$.

(e). Each users computes $U_S = 34$, $V_S = 3$, $W_S = 18$ and $R_S = h(3, m) = 8$ (let).

(f). $S_2$ recovers $f_S(u_{S_2}) = 3$ and computes $MS_{S_2} = 5$ and $s_2 = 22$.

(g). $S_4$ recovers $f_S(u_{S_4}) = 4$ and computes $MS_{S_4} = 21$ and $s_4 = 11$.

(i). $S_6$ recovers $f_S(u_{S_6}) = 16$ and computes $MS_{S_6} = 12$ and $s_6 = 16$.

(j). $S_7$ recovers $f_S(u_{S_7}) = 19$ and computes $MS_{S_7} = 19$ and $s_7 = 12$.

(k). CTC produces a group signature $S_S = 15$ and sends { 15 , 34 , 18 , *m*} to the designated combiner *DC* of organization *R* as signature of the group *S* for the message *m*.

*5.4. Signature verification by the organization R*

    Suppose five users $R_1$, $R_3$, $R_4$, $R_5$ and $R_6$ want to verify the signature, then

(a). R1 recovers $f_R(u_{R_1}) = 21$ and computes $MS_{R_1} = 19$.

(b). R3 recovers $f_R(u_{R_3}) = 21$ and computes $MS_{R_3} = 4$.

(c). R4 recovers $f_R(u_{R_4}) = 15$ and computes $MS_{R_4} = 11$.

(d). R5 recovers $f_R(u_{R_5}) = 12$ and computes $MS_{R_5} = 9$.

(e). R6 recovers $f_R(u_{R_6}) = 18$ and computes $MS_{R_6} = 10$.

(f). *DC* computes the value $R_R = 3$ and then recovers $R_S = h(3, m) = 8$.

(g). *DC* checks the following congruence for a valid signature $2^{15} \stackrel{?}{\equiv} 3 \cdot 27^8 \bmod 45$.

This congruence holds so $\{15, 34, 18, m\}$ is a valid signature on the message $m$.

## 6. Remarks

In this paper, we have proposed a **threshold signature scheme with threshold verification.** To obtain this construction, we have used the ElGamal public key cryptosystem and Schnorr's signature scheme. The security of this cryptosystem is based on the discrete log problem. The signature generation is done by certain designated sub – groups of signers and the verification is done by certain designated sub-groups of the group of the receivers. Here designated sub – groups are characterized by threshold values. The threshold value can be different for signature generation and for signature verification. Until *t* (the threshold value of the group of senders) individuals act in collusion they will get no information about the group secret key. Similarly, until *k* (the threshold value of the group of recipients) individuals act in collusion they will get no information about the group secret key. The group public parameters *p, q, g* and a collision free one-way hash function *h* is same for both the organizations.

In any case of dispute between the group *S* and *R*, the CTC keeps the records of signatures and plays the role of a trusted judge. Since the CTC can checks the validity of the signature so when any third party needs the signature verification, the CTC convince the third party about the facts.

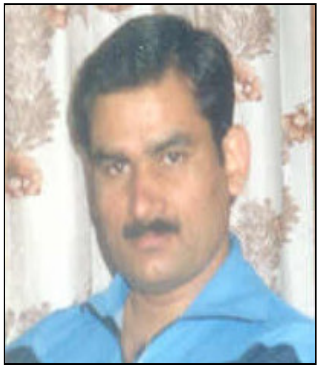

**Manoj Kumar** received the B.Sc. degree in mathematics from Meerut University Meerut, in 1993; the M. Sc. in Mathematics (Goldmedalist) from C.C.S.University Meerut, in 1995; the M.Phil. (Goldmedalist) in *Cryptography*, from Dr. B.R.A. University Agra, in 1996; submitted the Ph.D. thesis in *Cryptography*, in 2003. He also taught applied Mathematics at DAV College, Muzaffarnagar, India from Sep, 1999 to March, 2001; at S.D. College of Engineering & Technology, Muzaffarnagar, and U.P., India from March, 2001 to Nov, 2001; at Hindustan College of Science & Technology, Farah, Mathura, continue since Nov, 2001. He also qualified the *National Eligibility Test* (NET), conducted by *Council of Scientific and Industrial Research* (CSIR), New Delhi- India, in 2000. He is a member of Indian Mathematical Society, Indian Society of Mathematics and Mathematical Science, Ramanujan Mathematical society, and Cryptography Research Society of India. His current research interests include Cryptography, Numerical analysis, Pure and Applied Mathematics.